\documentclass[12pt]{article}
\usepackage{chngcntr}
\counterwithout{figure}{section}
\counterwithout{table}{section}
\usepackage{amssymb, amsmath}
\usepackage[colorlinks=true, allcolors=blue]{hyperref}
\usepackage{multirow}
\usepackage{float}
\usepackage{graphicx}
\usepackage{array}
\usepackage{booktabs}
\usepackage[top=1in, bottom=1in, left=1in, right=1in]{geometry}
\usepackage{xcolor}
\usepackage{setspace}  % Add this line
\usepackage{csquotes}  % Add csquotes package
\usepackage{natbib} 
\title{Formulating the Proxy Pattern-Mixture Model as a Selection Model to Assist with Sensitivity Analysis}
\author{Seth Adarkwah Yiadom\thanks{Ph.D. Student, Interdisciplinary Ph.D. Program in Biostatistics, The Ohio State University, Columbus, OH 43220. Email: }, 
Rebecca Andridge\thanks{Professor, Division of Biostatistics, The Ohio State University College of Public Health, 1841 Neil Ave., Columbus, OH 43220. Email: andridge.1@osu.edu}}
\date{}

\begin{document}
\maketitle

\singlespacing
\begin{abstract}
Proxy pattern-mixture models (PPMM) have previously been proposed as a model-based framework for assessing the potential for nonignorable nonresponse in sample surveys and nonignorable selection in nonprobability samples. One defining feature of the PPMM is the single sensitivity parameter, $\phi$, that ranges from 0 to 1 and governs the degree of departure from ignorability. While this sensitivity parameter is attractive in its simplicity, it may also be of interest to describe departures from ignorability in terms of how the odds of response (or selection) depend on the outcome being measured. In this paper, we re-express the PPMM as a selection model, using the known relationship between pattern-mixture models and selection models, in order to better understand the underlying assumptions of the PPMM and the implied effect of the outcome on nonresponse. The selection model that corresponds to the PPMM is a quadratic function of the survey outcome and proxy variable, and the magnitude of the effect depends on the value of the sensitivity parameter, $\phi$ (missingness/selection mechanism), the differences in the proxy means and standard deviations for the respondent and nonrespondent populations, and the strength of the proxy, $\rho^{(1)}$. Large values of $\phi$ (beyond $0.5$) often result in unrealistic selection mechanisms, and the corresponding selection model can be used to establish more realistic bounds on  $\phi$. We illustrate the results using data from the U.S. Census Household Pulse Survey.
\end{abstract}

\newpage

\doublespacing

\section{Introduction}
The proxy pattern mixture model (PPMM) is a method for assessing potentially nonignorable nonresponse bias for the mean of a survey variable subject to nonresponse, when there is a set of covariates observed for nonrespondents and respondents \citep{andridge2011proxy}. It is a model-based method to create adjusted estimators for situations where nonresponse could be nonignorable and uses a sensitivity analysis to assess how much inference is affected by nonignorable nonresponse. The PPMM provides a simple framework for addressing issues arising from missing data due to nonresponse or issues of selection bias. In the context of missing data, \cite{andridge2011proxy, andridge2020proxy} proposed the PPMM as a measure of the potential impact of nonresponse on survey estimates for both continuous and binary outcome variables. \cite{little2020measures} used the PPMM as the basis of indices that measure the degree of potential sampling bias arising from the use of nonprobability samples. \cite{andridge2019indices} extended this approach to indices of potential nonignorable selection bias for estimates of population proportions. The PPMM and indices derived from the PPMM have subsequently been shown to effectively capture nonignorable nonresponse and/or selection bias in different applications, including state-level pre-election polls in the U.S. and Great Britain \citep{west2023evaluating, jackson2023can}, for a variety of outcome measures from the German General Social Survey \citep{hammon2024validating}, and for estimates of COVID-19 vaccine from large-scale internet-based surveys in the U.S. \citep{andridge2024using}.

The PPMM, unlike other weighting and imputation methods for adjusting for
survey nonresponse, does not assume the data are missing at random (MAR) \citep{andridge2011proxy, andridge2020proxy}. Rather, it utilizes a single sensitivity parameter ($\phi$) ranging between 0 and 1 that controls the deviation from MAR, where $\phi=0$ corresponds to MAR and $\phi=1$ corresponds to missingness depending only on the outcome of interest that is subject to missingness. \cite{little2020measures} recommended the use of a central value of $\phi =0.5$, if a single point estimate is desired. The key question is, how ``realistic" is a value of $\phi=0.5$ (or other values)? The goal of this paper is to connect the PPMM to a selection model framework to help judge the magnitude of nonignorable missingness or selection. Selection models are an alternative modeling method that model the effect of survey outcome, $Y$ on missingness or selection, and can thus often be more easily interpreted. In Section 2, we briefly review the general framework of the PPMM. In Section 3, we derive the selection model that corresponds to the PPMM. Section 4 applies the method to data from the Household Pulse Survey, and Section 5 presents discussion, including extensions of the proposed method.

\section{Background: The Proxy Pattern-Mixture Model}

 \cite{andridge2011proxy} first introduced the PPMM for assessing the potential for nonresponse bias for continuous outcomes in surveys. Since then, there have been several extensions of the methods to incorporate binary and skewed survey outcome cases \citep{andridge2019indices, andridge2020proxy, andridge2015assessing, andridge2015using} and to increase robustness \citep{yang_little_2021}. This section contains a brief overview of the PPMM in the context of estimating means; we refer interested readers to the aforementioned references for more details. 

\subsection{PPMM General Framework for Continuous Survey Outcome}
We consider a sample size of $n$ units drawn randomly from an infinite population. For the $i^{th}$ unit in the sample, let $Y_i$ denote the value of the continuous survey outcome variable and $Z_i = (Z_{i1}, Z_{i2},...,Z_{ip})$ denote the observed values of $p$ covariates. We do not observe the responses of all the $n$ subjects in the sample, rather, we only observe the responses of $r$ out of the $n$ sampled units. Thus, the observed data consists of $(Y_i, Z_i)$ for $i=1,2,.., r$ and $Z_i$ for $i=r+1,...,n$. Assessing and correcting for the nonresponse bias in the survey outcome mean is of primary interest.

Let $R$ denote the response  indicator, where $R=0$ when $Y$ is missing and $R=1$ when $Y$ is observed.  As described by \cite{andridge2011proxy}, we reduce the covariates $Z$ to a single proxy variable $X$ that is a linear combination of the $p$ covariates $Z$ by regressing $Y$ on $Z$ using the data for the respondent group, i.e., $E(Y|Z, R=1)=\alpha_0 +\alpha_1 Z$. The proxy variable, $X$, is obtained as the predicted values of the survey outcome $Y$ from the regression model, and is available for both non-respondents and respondents. 

The joint distribution of $[Y, X, R]$ is given by:

\begin{align}\label{jointdist}
\begin{pmatrix}\left. \begin{array}{c} Y \\[0.3em]
 X \\[0.3em]  \end{array} \right| R = r
\end{pmatrix} &\sim  N
\begin{bmatrix}
\begin{pmatrix}
\mu_y^{(r)}\\
\mu_x^{(r)}
\end{pmatrix}\!\!,&
\begin{pmatrix}
{\sigma_y^{(r)}}^2 &  \sigma_{xy}^{(r)} \\
\sigma_{xy}^{(r)}  & {\sigma_x^{(r)}}^{2} 
\end{pmatrix}
\end{bmatrix} \\ \nonumber \\
R & \sim Bernoulli(\pi) \nonumber
\end{align}

\noindent where $\pi$ is the response rate for the outcome $Y$ (i.e., $r/n$). The parameters $\mu_y^{(0)}, {\sigma_y^{(0)}}^2$, and $\sigma_{xy}^{(0)}$ for non-respondents are unidentifiable without any further assumptions. To identify these parameters from the model, an identifying restriction is imposed on the parameters \citep{andridge2011proxy, andridge2020proxy, little2020measures}: 
\begin{equation}\label{Idassump}
    P(R=1 | Y, X) = f\{(1-\phi)X^{*}+\phi Y, ~V\}.
\end{equation}
Here, $f$ is an unspecified function,  $X^{*}= X(\sigma_y^{(1)} / \sigma_x^{(1)})$ is a rescaled (for mathematical convenience) version of the proxy $X$, and $V$ is a set of other covariates that are independent of $Y$ and $X$ for units in the respondent sample but that may be related to response (i.e., related to $R$). The parameter $\phi$ is a sensitivity parameter -- there is no information in the data that can be used to estimate it -- that is varied to account for a range of assumptions on the missingness mechanism. The relative amount of missingness due to $X$ and to $Y$ is effectively described by the values of $\phi$, which ranges from $0$ to $1$. If $\phi=0$, it means that the missingness solely depends on $X$, not $Y$, i.e., the data is missing at random (MAR). On the extreme side, if $\phi=1$, then missingness solely depends on $Y$, not $X$, i.e., the data is ``extremely" missing not at random (MNAR). Values of $\phi$ between 0 and 1 correspond to nonignorable missingness mechanisms where missingness depends in part on the outcome $Y$.

As shown in \citet{andridge2011proxy}, imposing the assumption given in Equation (\ref{Idassump}) results in identification of $\{\mu_y^{(0)}, {\sigma_y^{(0)}}^2, \sigma_{xy}^{(0)}\}$ for a given value of $\phi$:
\begin{equation}\label{UnId1}
    \mu_y^{(0)} = \mu_y^{(1)} + \dfrac{\sigma_y^{(1)}}{\sigma_x^{(1)}}\left(\dfrac{\phi + (1-\phi)\rho^{(1)}}{\phi \rho^{(1)}+(1-\phi)} \right) \left(\mu_x^{(0)} - \mu_x^{(1)} \right)
\end{equation}

\begin{equation}\label{UnId2}
    {\sigma_y^{(0)}}^{2} = {\sigma_y^{(1)}}^{2}+ \dfrac{ {\sigma_y^{(1)}}^{2}}{{\sigma_x^{(1)}}^2} \left(\dfrac{\phi + (1-\phi)\rho^{(1)}}{\phi \rho^{(1)}+(1-\phi)} \right)^2 \left({\sigma_x^{(0)}}^{2} - {\sigma_x^{(1)}}^{2} \right)
\end{equation}

\begin{equation}\label{UnId3}
    \sigma_{xy}^{(0)} = \sigma_{xy}^{(1)}+ \dfrac{ \sigma_y^{(1)}}{\sigma_x^{(1)}}\left(\dfrac{\phi + (1-\phi)\rho^{(1)}}{\phi \rho^{(1)}+(1-\phi)} \right) \left({\sigma_x^{(0)}}^{2} - {\sigma_x^{(1)}}^{2} \right)
\end{equation}
\noindent Here $\rho^{(1)} = \frac{\sigma_{xy}^{(1)}}{\sigma_y^{(1)}\sigma_x^{(1)}}$ is the correlation between the outcome $Y$ and proxy $X$ in the respondent sample and is referred to as the ``strength" of the proxy. The overall mean for $Y$ is then the weighted average of the respondent and nonrespondent means, i.e., $\mu_y=\pi \mu_y^{(1)} + (1-\pi)\mu_y^{(0)}$. By either fixing $\phi$ at a single value or putting a prior on $\phi$ (e.g., Uniform(0,1)), one can estimate the mean of $Y$ under the PPMM and estimate the amount of bias in the respondent mean.

\section{The PPMM as a Selection Model}

Pattern-mixture models and selection models are two contrasting approaches when dealing with incomplete data and nonignorable missing-data mechanisms \citep{glynn2013selection}. Both approaches rely on assumptions that cannot be empirically verified and require additional constraints to determine the parameters. \cite{kaciroti2014bayesian} showed the equivalence between these two approaches for several common outcome types. By applying their results to the PPMM, we show how the PPMM can be expressed as a selection model to assist with the sensitivity analysis.

\subsection{PPMM for Continuous Y as a Selection Model}

Following the results by \cite{kaciroti2014bayesian}, the full selection model corresponding with the PPMM given by Equations (\ref{jointdist}) and (\ref{Idassump}) is:

\begin{equation}\label{FullSM}
\text{logit}(P(R=0|X,Y)) = \lambda_0 + \lambda_1 x + \lambda_2 x^2 + \lambda_3 y + \lambda_4 xy + \lambda_5 y^2
\end{equation}
where 
\begin{align*}
    \lambda_0 & = \text{log}\left(\dfrac{1-\pi}{\pi} \right) + \dfrac{{\mu_x^{(1)}}^2}{2{\sigma_x^{(1)}}^2} - \dfrac{{\mu_x^{(0)}}^2}{2{\sigma_x^{(0)}}^2} + \dfrac{1}{2}\text{log} \left(\dfrac{{\sigma_x^{(1)}}^2}{{\sigma_x^{(0)}}^2} \right) + \dfrac{{\beta^{(1)}}^2}{2{\sigma_{y|x}^{(1)}}^2} - \dfrac{{\beta^{(0)}}^2}{2{\sigma_{y|x}^{(0)}}^2} + \dfrac{1}{2}\text{log} \left(\dfrac{{\sigma_{y|x}^{(0)}}^2}{{\sigma_{y|x}^{(1)}}^2} \right) \\
    \lambda_1 & =  \dfrac{\mu_x^{(0)}}{{\sigma_x^{(0)}}^2} -  \dfrac{\mu_x^{(1)}}{{\sigma_x^{(1)}}^2} + \dfrac{\beta^{(1)} \alpha^{(1)}}{{\sigma_{y|x}^{(1)}}^2} - \dfrac{\beta^{(0)} \alpha^{(0)}}{{\sigma_{y|x}^{(0)}}^2} \\
    \lambda_2 & =  \dfrac{1}{2{\sigma_x^{(1)}}^2} - \dfrac{1}{2{\sigma_x^{(0)}}^2} +  \dfrac{{\alpha^{(1)}}^2}{2{\sigma_{y|x}^{(1)}}^2} - \dfrac{{\alpha^{(0)}}^2}{2{\sigma_{y|x}^{(0)}}^2} \\
    \lambda_3 &=   \dfrac{\beta^{(0)}}{{\sigma_{y|x}^{(0)}}^2} - \dfrac{\beta^{(1)}}{{\sigma_{y|x}^{(1)}}^2} \\
    \lambda_4 & = \dfrac{\alpha^{(0)}}{{\sigma_{y|x}^{(0)}}^2} - \dfrac{\alpha^{(1)}}{{\sigma_{y|x}^{(1)}}^2} \\
    \lambda_5 & = \dfrac{1}{2{\sigma_{y|x}^{(1)}}^2} -\dfrac{1}{2{\sigma_{y|x}^{(0)}}^2}.
\end{align*}
Here, $\alpha^{(r)}$, $\beta^{(r)}$, and ${\sigma_{y|x}^{(r)}}^2$ are the parameters of the conditional distribution of $Y$ given $X$ and $R$:
\begin{align*}
    \beta^{(r)} & =\mu_y^{(r)} - \dfrac{\sigma_{xy}^{(r)}}{{\sigma_x^{(r)}}^2}\mu_x^{(r)}\\
    \alpha^{(r)} & =\dfrac{\sigma_y^{(r)}}{\sigma_x^{(r)}}\\
    {\sigma_{y|x}^{(r)}}^2 & =\left(1-{\rho^{(r)}}^2 \right){\sigma_y^{(r)}}^2   ={\sigma_y^{(r)}}^2 - \dfrac{{\sigma_{xy}^{(r)}}^2}{ {\sigma_{x}^{(r)}}^2}.
\end{align*}

\noindent The derivation of this result is in the Appendix (Section \ref{sect:appendix}). 

When the respondent and nonrespondent proxy variances are equal, i.e., ${\sigma_x^{(0)}}^2 = {\sigma_x^{(1)}}^2$, then by \eqref{UnId2} and \eqref{UnId3} we have ${\sigma_y^{(0)}}^2 = {\sigma_y^{(1)}}^2$ and ${\sigma_{xy}^{(0)}}^2 = {\sigma_{xy}^{(1)}}^2$. As a result, $\lambda_2 = \lambda_4 = \lambda_5 = 0$ and the selection model is linear in $Y$ and $X$ with no interaction term. In this case, a single odds ratio describes the relationship between the outcome, $Y$, and the probability of nonresponse. However, when the respondent proxy variance is not equal to the nonrespondent variance, this relationship depends on both $Y$ and $X$ due to the quadratic term $(\lambda_5)$ and interaction term $(\lambda_4)$.

\subsection{Visualization of the Corresponding Selection Model}

The selection model \eqref{FullSM} that corresponds to the PPMM can be difficult to understand due to the quadratic and interaction terms. Visualizing the results in terms of odds ratios can facilitate its interpretation. We computed the odds ratio of nonresponse for a $1$ unit increase in $Y$ (based on \eqref{FullSM}) as a function of the sensitivity parameter $\phi$ for various combinations of the PPMM parameters. We considered a fixed response rate of $\pi = 0.75$. For respondents, we fixed the mean and variance of the proxy variable $X$ as well as the outcome $Y$ at $1$, but varied $\rho^{(1)}$ at $\{0.2, 0.5, 0.8\}$ to vary the strength of the correlation between the proxy and the survey outcome variable. We varied the nonrespondent mean and variance of the proxy at $\mu_x^{(0)} = \{0.8, 1.2\}$ and ${\sigma_x^{(0)}}^2 = \{0.9, 1, 1.1\}$ respectively. We considered all the possible combinations of the varied parameters to obtain 18 different mechanisms, i.e., a $3 \times 2 \times 3$ factorial design.

\begin{figure}[H]
\centerline{\includegraphics[scale=1]{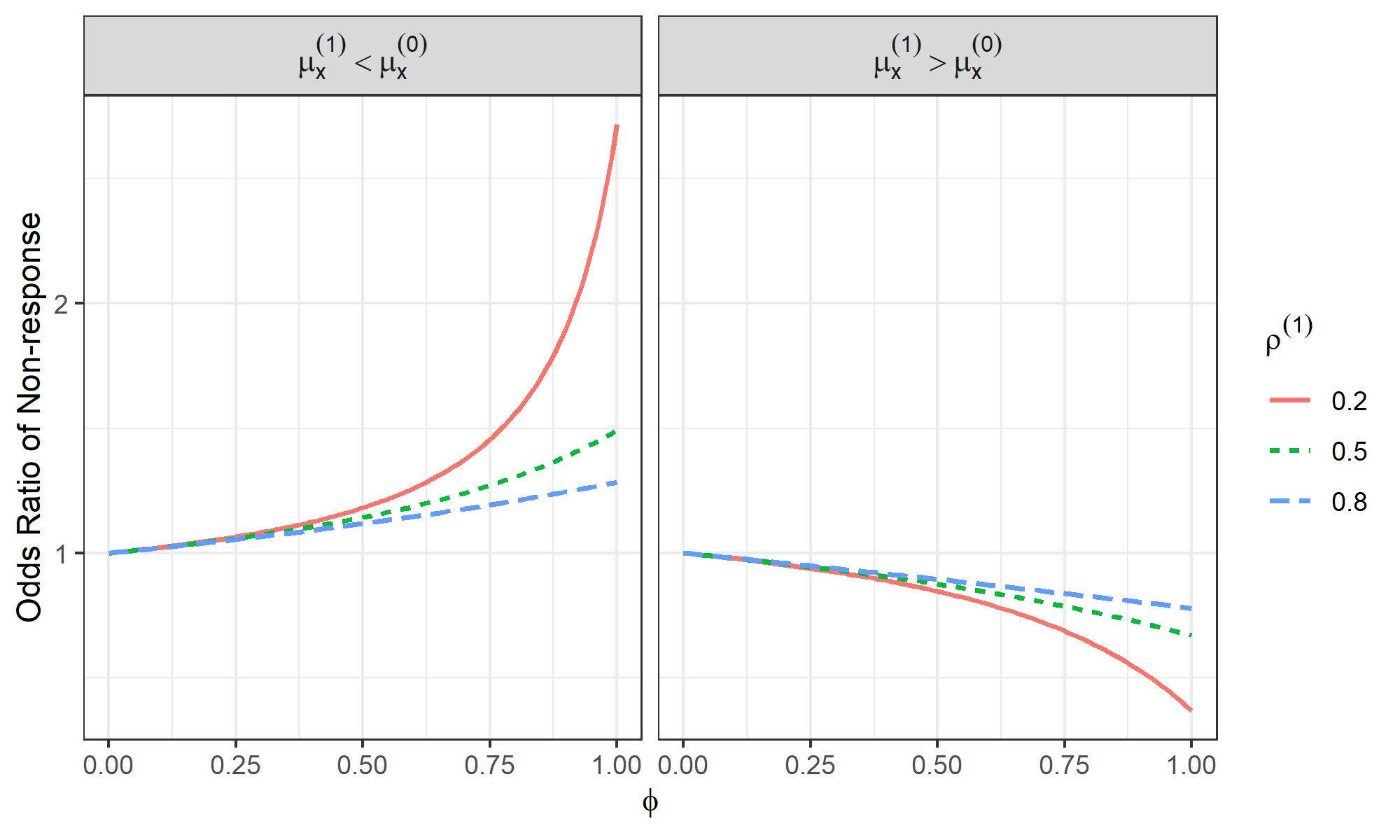}}
\caption{Odds ratio of nonresponse for a 1 standard deviation increase in $Y$ as a function of $\phi$ when ${\sigma_x^{(1)}}^2 = {\sigma_x^{(0)}}^2$. The proxy $X$ was fixed at its overall mean.}
\label{fig:OR_mech_7_12}
\end{figure}

Figure \ref{fig:OR_mech_7_12} shows the selection model odds ratio for a $1$ unit increase in $Y$ as a function of the sensitivity parameter $\phi$ for the cases when the respondent and nonrespondent proxy variances are equal. When $\phi=0$, nonresponse is ignorable, so the corresponding odds ratio for $Y$ is 1. The odds ratio moves away from 1 as $\phi$ increases, with the direction of the movement depending on whether the proxy mean is larger (left panel) or smaller (right panel) among nonrespondents. This makes sense -- larger values of $\phi$ correspond to missingness depending more strongly on $Y$. Importantly, for weaker proxies (smaller correlations), the odds ratio increases (decreases) at a more drastic rate. This illustrates a built-in assumption of the PPMM, that by definition weaker proxies will correspond to more severe missingness mechanisms (given similar proxy means and variances). However, even for the weakest proxy strength, the odds ratios are arguably not too unrealistic -- these are the odds ratio for a $1$ unit increase in $Y$, which is a one standard deviation increase, and the odds ratio only gets as large as approximately $2$ when $\phi =0.9$ (and similarly for small odds ratios).

\begin{figure}[th]
\centerline{\includegraphics[scale=1]{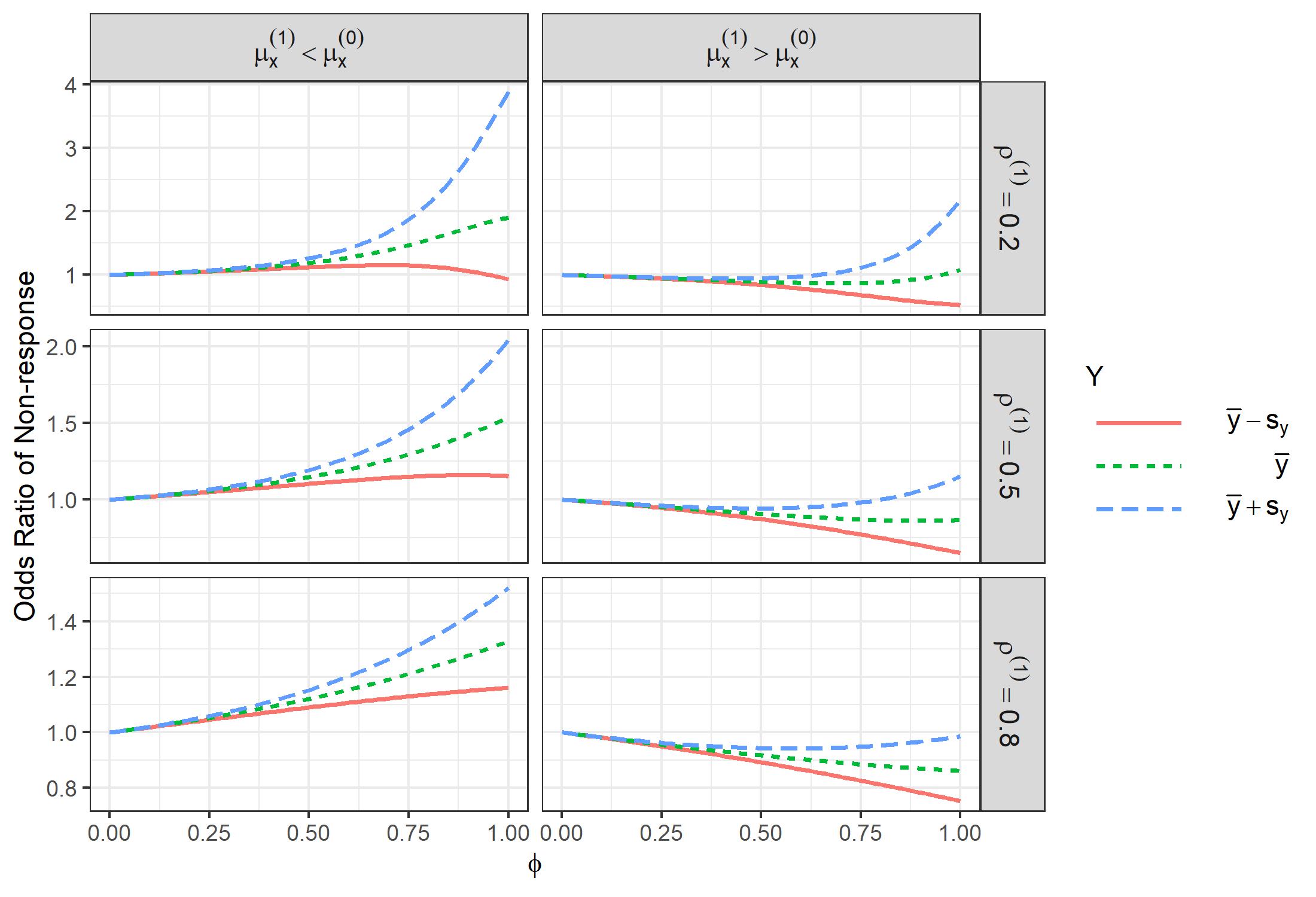}}
\caption{Odds ratio of nonresponse as a function of $\phi$ when ${\sigma_x^{(1)}}^2 < {\sigma_x^{(0)}}^2$ for different levels of $Y$: the respondent mean and 1 standard deviation below and above the respondent mean.  The proxy variable was fixed at the overall mean of $X$.}
\label{fig:OR_mech_13_18}
\end{figure}

Figure \ref{fig:OR_mech_13_18} shows the selection model odds ratios for the cases when the respondent proxy variance is smaller than the nonrespondent proxy variance. In this case, the odds ratio for $Y$ depends on $X$ and on $Y$ itself due to the quadratic and interaction terms. Thus we computed odds ratios at three different values of $Y$ (the respondent mean of $Y$ plus/minus one standard deviation) and fixed the proxy $X$ at its overall mean.

We see a similar pattern as in Figure \ref{fig:OR_mech_7_12} across all facets, with the odds ratio for nonresponse tending to increase in strength (move away from 1) as $\phi$ increases, with the rate of increment depending on the strength of the correlation. There is some evidence of potentially implausibly large odds ratios when $\phi$ approaches 1, particularly for weak proxies. In addition, in some instances the quadratic relationship causes some undesirable results for large $\phi$ values. For example, in the upper left panel of Figure \ref{fig:OR_mech_13_18}, when $Y$ is one standard deviation below its mean (red line), the odds ratio initially increases as a function of $\phi$ but then starts to decrease, eventually crossing one when $\phi$ is approximately 0.9. This implies that, for $\phi$ close to 1, increasing $Y$ is associated with \textit{decreased} odds of nonresponse -- the opposite of what is expected and what the rest of the plot shows. This information could inform a more tight bound on $\phi$ for a sensitivity analysis, excluding the $\phi$ values that lead to this type of selection mechanism.

When the respondent proxy variance is greater than the nonrespondent variance, similar results are obtained (Supplemental Figure A1). Odds ratios are slightly larger and the direction of curvature is flipped between the smaller and larger values of $Y$.

\section{Application to Census HPS}
We applied our results to a real data set, the U.S. Census Bureau’s Household Pulse Survey (HPS) \citep{USCensus_HPS}, to illustrate how it could inform the PPMM sensitivity analysis. The HPS was designed to collect data on household experiences during the coronavirus pandemic in a timely and efficient manner. It has evolved to include information on other emerging social and economic challenges affecting households across the country. The HPS is an online self-response survey, where questions are asked of individuals who are 18 years and above. The survey started in April 2020 and has since been collected in phases with each phase having many waves \citep{USCensus_HPS}. For the purpose of our study, we use data from phase 3.1, wave 29 (week 29) which was collected from April 28 to May 10, 2021. 

We used the Patient Health Questionnaire for Depression and Anxiety (PHQ-4) score as the outcome of interest, $Y$. The PHQ-4 is a 4-item ultra-brief depression score for detecting both anxiety and depressive disorders \citep{kroenke2009ultra, christodoulaki2022validation}. The PHQ-4 score is based on responses to the following four questions: frequency of anxiety over the previous 7 days, frequency of worry over the previous 7 days, frequency of having little interest in things over previous 7 days, and frequency of feeling depressed over previous 7 days. Responses to each question are scored as 0 (“not at all”), 1 (“several days”), 2 (“more than half the days”), and 3 (“nearly every day”) and the PHQ-4 score is the sum of responses to the four items. Therefore, the total score on this composite measure ranges from 0 to 12 with higher scores indicating worse mental health. We considered the overall PHQ-4 score to be missing if any of the items were missing, resulting in missing PHQ-4 scores for $17.7\%$ of the sample. A plausible explanation for missing PHQ-4 scores is that individuals with worse mental health are less likely to answer these questions, thus leading to a nonignorable nonresponse mechanism.

\begin{table}[ht]
\caption{Covariates used to create the proxy for the application to the Census Household Pulse Survey}
\label{tab:VariableTable}
\begin{tabular}{ll}
\toprule
Variable & Description/Levels \\ \midrule
Age & Age of respondent (years) \\ [5pt]
Gender & 1) Male \\
 & 2) Female \\ [5pt]
Race & 1) White, alone \\
& 2) Black, alone \\
& 3) Asian, alone  \\
& 4) Any other race alone, or race in combination \\ [5pt]
Ethnicity & 1) No, not of Hispanic, Latino, or Spanish origin \\ 
& 2) Yes, of Hispanic, Latino, or Spanish origin \\ [5pt]
Education Level & 1) Less than high school \\
& 2) Some high school\\
& 3) High school graduate or equivalent\\
& 4) Some college, but degree not received or is in progress\\
& 5) Associate’s degree\\
& 6) Bachelor's degree\\ 
& 7) Graduate degree\\ [5pt]
Geographic (Census) Region & 1) Northeast \\
& 2) South\\ 
& 3) Midwest\\
& 4) West \\
\bottomrule
\end{tabular}
\end{table}

We created the proxy for the PHQ-4 score using a linear regression model with the main effects of all variables in Table \ref{tab:VariableTable} as predictors. Note that while the HPS does have sampling weights, our analysis focuses on understanding the potential for a nonignorable response mechanism for the PHQ-4 score among individuals who participated in the survey, and thus ignored these weights. Therefore, mean estimates should not be considered generalizable to the U.S. population. The resulting correlation between the proxy ($X$) and the outcome variable (PHQ-4) among respondents was $\hat{\rho}^{(1)}=0.28$, and the respondent proxy mean ($\hat{\mu}_x^{(1)}=2.75$) and variance ($\hat{\sigma}_{xx}^{(1)}=0.96$) were smaller than the corresponding values for nonrespondents ($\hat{\mu}_x^{(0)}=3.04, \hat{\sigma}_{xx}^{(0)}=1.02$).

\subsection{Results}

\begin{figure}[th]
\centerline{\includegraphics[scale=1]{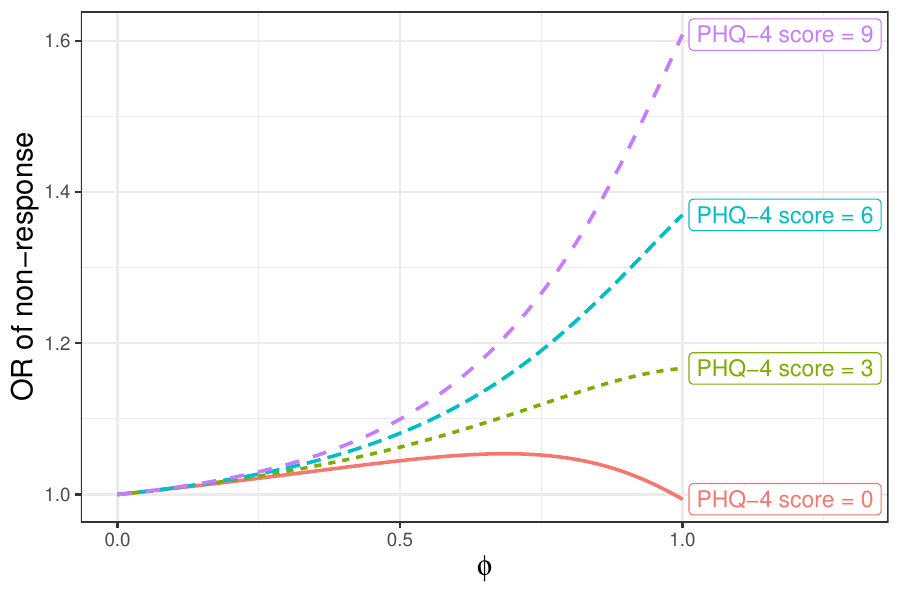}}
\caption{Odds ratio of non-response (for a +1 unit increase in PHQ-4 score) as a function of $\phi$ for different levels of the outcome variable (PHQ-4 score) with the proxy variable $X$ fixed at its overall mean.}
\label{fig:PHBCont_or}
\end{figure}

Figure \ref{fig:PHBCont_or} shows the odds ratio of non-response for $Y$ (PHQ-4 score) as a function of $\phi$ for various values of $Y$, with the proxy $X$ fixed at its mean. The plot looks similar to the top-left panel of Figure \ref{fig:OR_mech_13_18}. The odds ratio capturing the effect of the outcome, PHQ-4 score, on nonresponse tends to increase above 1 in general as $\phi=0$ increases, indicating that as $\phi$ increases, the strength of the relationship between PHQ-4 score and nonresponse increases, with higher PHQ-4 scores associated with higher odds of nonresponse. However, the rate of increment depends on the value of PHQ-4 score since the selection model is quadratic in $Y$. The odds ratio increases more rapidly as a function of $\phi$ for large PHQ-4 score values compared to smaller values. At the lowest PHQ-4 score, the odds ratio initially increases but then starts to decrease at about $\phi=0.75$, eventually dipping below 1 when $\phi$ is close to 1. This suggests that values of $\phi$ close to 1 are unrealistic for this outcome.

\begin{figure}[th]
\centerline{\includegraphics[scale=1]{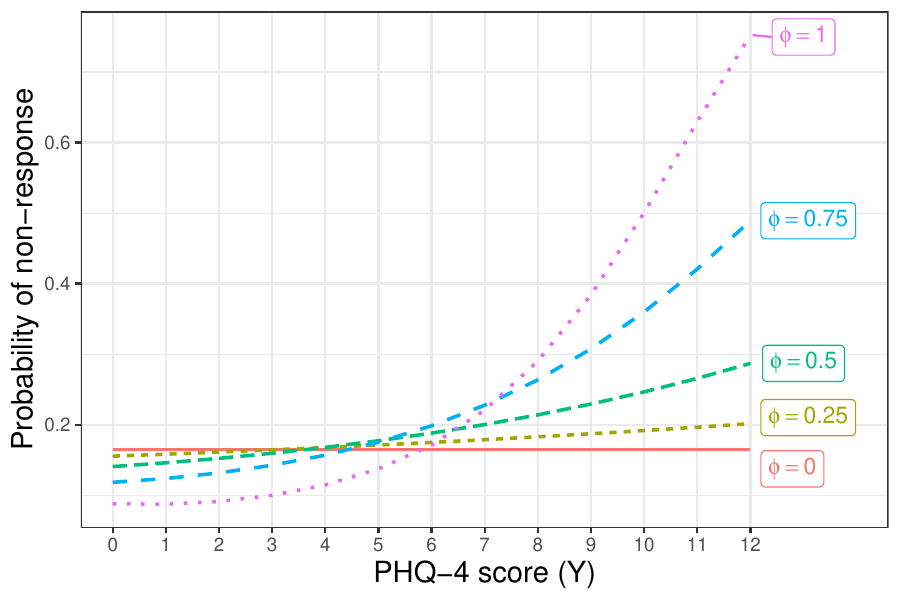}}
\caption{\textit{Probability of non-response  as a function of the outcome, PHQ-4, for different levels of the sensitivity parameter $\phi$, with the proxy variable $X$ fixed at its mean in the full data.}}
\label{fig:PHBCont_prob}
\end{figure}

We can also visualize the probability of a missing PHQ-4 score under the PPMM as a function of PHQ-4; these probabilities come directly from the logistic selection model. Figure \ref{fig:PHBCont_prob} shows the probability of missing PHQ-4 score as a function of PHQ-4 score, for different values of the PPMM sensitivity parameter $\phi$. At $\phi=0$, the missingness mechanism is ignorable, and thus the line is flat; for values of $\phi>0$ the probability of nonresponse increases as the PHQ-4 score increases. Larger values of $\phi$ are ``more non-ignorable" and thus have steeper curves. For $\phi=1$ the implied probability of nonresponse becomes very large, approximately 0.5 for a PHQ-4 score of 10 and 0.75 for a PHQ-4 score of 12. These probabilities are quite extreme (e.g., 75\% of individuals with a PHQ-4 score of 12 are missing the outcome), suggesting that this large a $\phi$ value may be unrealistic. In fact, for $\phi=0.75$ we also see fairly large probabilities, just shy of 0.5 for the largest PHQ-4 scores. This suggests that $\phi=0.75$ might be a reasonable upper bound to use in a sensitivity analysis using the PPMM.

\section{Discussion}
In this paper, we have explored the equivalence and application of proxy pattern-mixture models (PPMM) in a selection model framework to help judge the magnitude of nonignorable missingness or selection under the PPMM. The integration of these two models helps translate the PPMM sensitivity parameter $\phi$ into a potentially easier to interpret odds ratio, and can potentially facilitate a sensitivity analysis by informing tighter bounds on $\phi$. The selection model that corresponds to the proxy-pattern mixture model is a quadratic function of the survey outcome ($Y$) and proxy variable ($X$). However, the magnitude of the quadratic effect depends on the strength of the proxy (i.e., how well the covariates predict the survey outcome among respondents) with more severe quadratic effecits for weaker proxies. Having access to strong predictors of the survey outcome of interest has previously been suggested as important for the PPMM \citep[e.g.,][]{little2020measures} and our analysis reveals that in fact the effect of $Y$ on response can sometimes be implausible under the PPMM with weaker covariates (e.g., effects changing direction). 

Regardless of the strength of the proxy, our results show that the odds ratio for $Y$ becomes exaggerated for $\phi$ values close to one. This can lead to unrealistic or more extreme estimates of the nonresponse mechanism, highlighting the need for careful selection and justification of the choice of $\phi$ for a sensitivity analysis. This selection model formulation suggests we might want to place an upper bound on the sensitivity parameter that is smaller than $1$, as $\phi = 1$ may be too extreme in some scenarios.

A limitation of our visualizations is that we fixed the proxy $X$ at its mean to reduce complexity. This simplification may not fully capture the variability and potential interactions present in real-world scenarios, and future research should explore the impact of allowing the proxy variable to vary. By doing so, we can gain a more comprehensive understanding of the dynamics at play. In addition, we did not use a Bayesian approach to put a prior on the sensitivity parameter as has been previously suggested \cite{little2020measures, andridge2019indices}. Such an approach could have provided a posterior distribution for the odds ratio of $Y$.

One promising area for future research is the use of this selection model expression of the PPMM to obtain nonignorable nonresponse propensities for given values of $\phi$. This could enable the construction of inverse propensity weights under various degrees of nonignorable nonresponse. This may provide a straightforward way to estimate subgroup effects under the PPMM, which in its current form requires reapplying the method with subsets of data and, in the case of application of the PPMM to selection issues, separate population-level aggregate information on the subset, which might not be readily available. Our results therefore provide a basis for the development of nonignorable weights to adjust for potential biases introduced by the relationship between the survey outcome and nonresponse.

%%%%%%%%%%%%%%%% APPENDIX %%%%%%%%%%%%%%%
\newpage
\section{Appendix} \label{sect:appendix}

Following the results by  \cite{kaciroti2014bayesian} in the case of no observed covariate, we assume that the distribution of the observed $y$ is a member of the exponential family: 

\begin{equation}\label{ExpFam1}
f_r \left(y|\theta^{(r)}, \phi^{(r)}\right) = exp \Bigl\{ \dfrac{\theta^{(r)}y - b(\theta^{(r)})}{a(\phi^{(r)})} + c(y, \phi^{(r)}) \Bigr\}
\end{equation}
For normal outcome with no covariate, the proxy-pattern mixture model is $Y|R=r \sim N \left(\mu^{(r)}, {\sigma^{(r)}}^2\right)$ with its corresponding exponential family expression:

\begin{align}\label{ExpFam2}
    f_r\left(y|\mu^{(r)}, {\sigma^{(r)}}^2 \right) & = \dfrac{1}{\sqrt{2\pi {\sigma^{(r)}}^2}}exp \Bigl\{ \dfrac{-(y-\mu^{(r)})^2}{2{\sigma^{(r)}}^2} \Bigr\} \nonumber \\
    & = exp \biggl\{ \dfrac{\mu^{(r)}y - \frac{{\mu^{(r)}}^2}{2} } {{\sigma^{(r)}}^2} -\dfrac{y^2}{2{\sigma^{(r)}}^2} - \dfrac{1}{2}\text{log} \left(2\pi {\sigma^{(r)}}^2 \right) \biggr\} 
\end{align}
Comparing Equations (\ref{ExpFam1}) and (\ref{ExpFam2}), we deduce that $\theta^{(r)} = \mu^{(r)}$, $\phi^{(r)} ={\sigma^{(r)}}^2 $, $b(\theta^{(r)})=\frac{{\mu^{(r)}}^2}{2}$, $a(\phi^{(r)}={\sigma^{(r)}}^2$, and $c(y, \phi^{(r)})=-\frac{y^2}{2{\sigma^{(r)}}^2}-\frac{1}{2}\text{log} \left(2\pi {\sigma^{(r)}}^2 \right)$. This implies that $\frac{b(\theta^{(r)})}{a(\phi^{(r)})}=\frac{{\mu^{(r)}}^2}{2{\sigma^{(r)}}^2}$ and $\frac{\theta^{(r)}}{a(\phi^{(r)})}=\frac{\mu^{(r)}}{2{\sigma^{(r)}}^2}$. Using the results in Proposition 1, Equation 3 of Kaciroti and Raghunathan (2014), the corresponding selection model is:

\begin{equation}\label{SM}
    \text{logit}(P(R=0~|~y)) = \gamma_{0} + \gamma_{1} y + \gamma_{2} y^2
\end{equation}

where 
\begin{align*}
    \gamma_0 &= \text{log}\left(\dfrac{P(R=0)}{P(R=1)} \right) + \dfrac{{\mu^{(1)}}^2}{2{\sigma^{(1)}}^2} - \dfrac{{\mu^{(0)}}^2}{2{\sigma^{(0)}}^2} + \dfrac{1}{2}\text{log} \left(\dfrac{{\sigma^{(1)}}^2}{{\sigma^{(0)}}^2} \right) \\
    \gamma_1 &=  \dfrac{\mu^{(0)}}{{\sigma^{(0)}}^2} -  \dfrac{\mu^{(1)}}{{\sigma^{(1)}}^2} ~~~~~~\text{and}~~~~ \gamma_2 =  \dfrac{1}{2{\sigma^{(1)}}^2} - \dfrac{1}{2{\sigma^{(0)}}^2}.
\end{align*}
To incorporate covariate, $X$ into the selection model (Equation \ref{SM}), we use the conditional mean and conditional variance (also called residual variance) of $y$ in place of $\mu^{(r)}$ and ${\sigma^{(r)}}^2$. That is, we substitute $E(Y|X,R=r)$ and $V(Y|X, R=r)={\sigma_{y|x}^{(r)}}^2$ for $\mu^{(r)}$ and ${\sigma^{(r)}}^2$  respectively, where $E(Y|X,R=r)$ and $V(Y|X, R=r)$ are defined as

\begin{align}\label{UnId5}
    E(Y|X, R=r) & =  \mu_y^{(r)} + \dfrac{\sigma_y^{(r)}}{\sigma_x^{(r)}}\rho^{(r)}\left (X-\mu_x^{(r)} \right) \nonumber \\
    & = \mu_y^{(r)} - \dfrac{\sigma_{xy}^{(r)}}{{\sigma_x^{(r)}}^2}\mu_x^{(r)} + \dfrac{\sigma_y^{(r)}}{\sigma_x^{(r)}}X \nonumber \\
    & = \beta^{(r)} + \alpha^{(r)}X
\end{align}
\begin{align}\label{UnId4}
    V(Y|X, R=r)  = {\sigma_{y|x}^{(r)}}^2 & =\left(1-{\rho^{(r)}}^2 \right){\sigma_y^{(r)}}^2 \nonumber \\
    & ={\sigma_y^{(r)}}^2 - \dfrac{{\sigma_{xy}^{(r)}}^2}{ {\sigma_{x}^{(r)}}^2}
\end{align}
where $\beta^{(r)} =\mu_y^{(r)} - \dfrac{\sigma_{xy}^{(r)}}{{\sigma_x^{(r)}}^2}\mu_x^{(r)}$ and $\alpha^{(r)} =\dfrac{\sigma_y^{(r)}}{\sigma_x^{(r)}}$.

After incorporating $X$ into the selection model in Equation (\ref{SM}), we obtain:
\begin{align}\label{gam1}
    \gamma_1 y &= \left( \dfrac{\beta^{(0)}+\alpha^{(0)}x}{{\sigma_{y|x}^{(0)}}^2} - \dfrac{\beta^{(1)}+\alpha^{(1)}x}{{\sigma_{y|x}^{(1)}}^2} \right)y \nonumber \\
    & = \left( \dfrac{\beta^{(0)}}{{\sigma_{y|x}^{(0)}}^2} - \dfrac{\beta^{(1)}}{{\sigma_{y|x}^{(1)}}^2}  \right)y + \left( \dfrac{\alpha^{(0)}}{{\sigma_{y|x}^{(0)}}^2} - \dfrac{\alpha^{(1)}}{{\sigma_{y|x}^{(1)}}^2} \right)xy
\end{align}
 \begin{align}\label{gam2}
     \gamma_2 y^2 &= \left( \dfrac{1}{2{\sigma_{y|x}^{(1)}}^2} -\dfrac{1}{2{\sigma_{y|x}^{(0)}}^2} \right)y^2
 \end{align}
and 
\begin{align}\label{gam0}
 \gamma_0 &=   \text{log}\left(\dfrac{P(R=0|X)}{P(R=1|X)} \right) + \delta + \dfrac{1}{2}\text{log} \left(\dfrac{{\sigma_{y|x}^{(0)}}^2}{{\sigma_{y|x}^{(1)}}^2} \right)
\end{align}
where 
\begin{align}\label{delta}
    \delta & =  \dfrac{\left(\beta^{(1)}+\alpha^{(1)}x\right)^2}{2{\sigma_{y|x}^{(1)}}^2} -  \dfrac{\left(\beta^{(0)}+\alpha^{(0)}x\right)^2}{2{\sigma_{y|x}^{(0)}}^2}  \nonumber\\
    & =  \dfrac{{\beta^{(1)}}^2}{2{\sigma_{y|x}^{(1)}}^2} - \dfrac{{\beta^{(0)}}^2}{2{\sigma_{y|x}^{(0)}}^2} + \left( \dfrac{\beta^{(1)} \alpha^{(1)}}{{\sigma_{y|x}^{(1)}}^2} - \dfrac{\beta^{(0)} \alpha^{(0)}}{{\sigma_{y|x}^{(0)}}^2}  \right)x + \left( \dfrac{{\alpha^{(1)}}^2}{2{\sigma_{y|x}^{(1)}}^2} - \dfrac{{\alpha^{(0)}}^2}{2{\sigma_{y|x}^{(0)}}^2} \right)x^2.
\end{align}
To obtain $\text{log} \left( \dfrac{P(R=0|X)}{P(R=1|X)} \right) = \text{logit} (P(R=0|X))$ term in Equation (\ref{gam0}), we further assume that $X|R=r \sim N \left(\mu_x^{(r)}, {\sigma_x^{(r)}}^2 \right)$ and then use the results from the selection model with no covariate, similar to Equation (\ref{SM}): 
\begin{equation}\label{logit}
    \text{logit}(P(R=0~|~x)) = \tau_{0} + \tau_{1} x + \tau_{2} x^2
\end{equation}
where 
\begin{align*}
    \tau_0 &= \text{log}\left(\dfrac{P(R=0)}{P(R=1)} \right) + \dfrac{{\mu_x^{(1)}}^2}{2{\sigma_x^{(1)}}^2} - \dfrac{{\mu_x^{(0)}}^2}{2{\sigma_x^{(0)}}^2} + \dfrac{1}{2}\text{log} \left(\dfrac{{\sigma_x^{(1)}}^2}{{\sigma_x^{(0)}}^2} \right) \\
    \tau_1 &=  \dfrac{\mu_x^{(0)}}{{\sigma_x^{(0)}}^2} -  \dfrac{\mu_x^{(1)}}{{\sigma_x^{(1)}}^2} ~~~~~~\text{and}~~~~ \tau_2 =  \dfrac{1}{2{\sigma_x^{(1)}}^2} - \dfrac{1}{2{\sigma_x^{(0)}}^2}.
\end{align*}
When we substitute Equations (\ref{delta}) and (\ref{logit}) into Equation (\ref{gam0}), the $\gamma_0$ term becomes:
\begin{equation}\begin{split}\label{gamma0}
    \gamma_0 = &\text{log}\left(\dfrac{P(R=0)}{P(R=1)} \right) + \dfrac{{\mu_x^{(1)}}^2}{2{\sigma_x^{(1)}}^2} - \dfrac{{\mu_x^{(0)}}^2}{2{\sigma_x^{(0)}}^2} + \dfrac{1}{2}\text{log} \left(\dfrac{{\sigma_x^{(1)}}^2}{{\sigma_x^{(0)}}^2} \right) + \dfrac{{\beta^{(1)}}^2}{2{\sigma_{y|x}^{(1)}}^2} - \dfrac{{\beta^{(0)}}^2}{2{\sigma_{y|x}^{(0)}}^2} + \dfrac{1}{2}\text{log} \left(\dfrac{{\sigma_{y|x}^{(0)}}^2}{{\sigma_{y|x}^{(1)}}^2} \right) \\
    & + \left(\dfrac{\mu_x^{(0)}}{{\sigma_x^{(0)}}^2} -  \dfrac{\mu_x^{(1)}}{{\sigma_x^{(1)}}^2} + \dfrac{\beta^{(1)} \alpha^{(1)}}{{\sigma_{y|x}^{(1)}}^2} - \dfrac{\beta^{(0)} \alpha^{(0)}}{{\sigma_{y|x}^{(0)}}^2} \right) x + \left( \dfrac{1}{2{\sigma_x^{(1)}}^2} - \dfrac{1}{2{\sigma_x^{(0)}}^2} +  \dfrac{{\alpha^{(1)}}^2}{2{\sigma_{y|x}^{(1)}}^2} - \dfrac{{\alpha^{(0)}}^2}{2{\sigma_{y|x}^{(0)}}^2} \right) x^2.
\end{split}\end{equation}

%%%%%%%%%%%%%%%%%%%% REFERENCES %%%%%%%%%%%%%%%%%%%
\newpage
\bibliography{references}

\begin{thebibliography}{}

\bibitem[Andridge and Thompson, 2015a]{andridge2015assessing}
Andridge, R. and Thompson, K.~J. (2015a).
\newblock {Assessing nonresponse bias in a business survey: Proxy
  pattern-mixture analysis for skewed data}.
\newblock {\em The Annals of Applied Statistics}, 9(4):2237 -- 2265.

\bibitem[Andridge and Thompson, 2015b]{andridge2015using}
Andridge, R. and Thompson, K.~J. (2015b).
\newblock Using the fraction of missing information to identify auxiliary
  variables for imputation procedures via proxy pattern-mixture models.
\newblock {\em International Statistical Review}, 83(3):472--492.

\bibitem[Andridge, 2024]{andridge2024using}
Andridge, R.~R. (2024).
\newblock Using proxy pattern-mixture models to explain bias in estimates of
  covid-19 vaccine uptake from two large surveys.
\newblock {\em Journal of the Royal Statistical Society Series A: Statistics in
  Society}, 187(3):831--843.

\bibitem[Andridge and Little, 2011]{andridge2011proxy}
Andridge, R.~R. and Little, R.~J. (2011).
\newblock Proxy pattern-mixture analysis for survey nonresponse.
\newblock {\em Journal of Official Statistics}, 27(2):153.

\bibitem[Andridge and Little, 2020]{andridge2020proxy}
Andridge, R.~R. and Little, R.~J. (2020).
\newblock Proxy pattern-mixture analysis for a binary variable subject to
  nonresponse.
\newblock {\em Journal of Official Statistics}, 36(3):703--728.

\bibitem[Andridge et~al., 2019]{andridge2019indices}
Andridge, R.~R., West, B.~T., Little, R.~J., Boonstra, P.~S., and
  Alvarado-Leiton, F. (2019).
\newblock Indices of non-ignorable selection bias for proportions estimated
  from non-probability samples.
\newblock {\em Journal of the Royal Statistical Society Series C: Applied
  Statistics}, 68(5):1465--1483.

\bibitem[Christodoulaki et~al., 2022]{christodoulaki2022validation}
Christodoulaki, A., Baralou, V., Konstantakopoulos, G., and Touloumi, G.
  (2022).
\newblock Validation of the patient health questionnaire-4 (phq-4) to screen
  for depression and anxiety in the greek general population.
\newblock {\em Journal of psychosomatic research}, 160:110970.

\bibitem[Glynn et~al., 2013]{glynn2013selection}
Glynn, R.~J., Laird, N.~M., and Rubin, D.~B. (2013).
\newblock Selection modeling versus mixture modeling with nonignorable
  nonresponse.
\newblock In {\em Drawing inferences from self-selected samples}, pages
  115--142. Routledge.

\bibitem[Hammon and Zinn, 2024]{hammon2024validating}
Hammon, A. and Zinn, S. (2024).
\newblock Validating an index of selection bias for proportions in
  non-probability samples.
\newblock {\em International Statistical Review}.

\bibitem[Jackson, 2023]{jackson2023can}
Jackson, M. (2023).
\newblock Can new metrics help us get a handle on partisan nonresponse bias?
  evidence from state-level 2022 polling.
\newblock In {\em 78th Annual AAPOR Conference}. AAPOR.

\bibitem[Kaciroti and Raghunathan, 2014]{kaciroti2014bayesian}
Kaciroti, N.~A. and Raghunathan, T. (2014).
\newblock Bayesian sensitivity analysis of incomplete data: bridging
  pattern-mixture and selection models.
\newblock {\em Statistics in medicine}, 33(27):4841--4857.

\bibitem[Kroenke et~al., 2009]{kroenke2009ultra}
Kroenke, K., Spitzer, R.~L., Williams, J.~B., and L{\"o}we, B. (2009).
\newblock An ultra-brief screening scale for anxiety and depression: the
  phq--4.
\newblock {\em Psychosomatics}, 50(6):613--621.

\bibitem[Little et~al., 2020]{little2020measures}
Little, R.~J., West, B.~T., Boonstra, P.~S., and Hu, J. (2020).
\newblock Measures of the degree of departure from ignorable sample selection.
\newblock {\em Journal of survey statistics and methodology}, 8(5):932--964.

\bibitem[{U.S. Census Bureau}, 2024]{USCensus_HPS}
{U.S. Census Bureau} (2024).
\newblock The {H}ousehold {P}ulse {S}urvey.
\newblock Retrieved from
  \url{https://www.census.gov/data/experimental-data-products/household-pulse-survey.html}.

\bibitem[West and Andridge, 2023]{west2023evaluating}
West, B.~T. and Andridge, R.~R. (2023).
\newblock Evaluating pre-election polling estimates using a new measure of
  non-ignorable selection bias.
\newblock {\em Public Opinion Quarterly}, 87(SI):575--601.

\bibitem[Yang and Little, 2021]{yang_little_2021}
Yang, Y. and Little, R.~J. (2021).
\newblock Spline pattern-mixture models for missing data.
\newblock {\em Journal of Data Science}, 19(1):75--95.

\end{thebibliography}

%%%%%%%%%%%%%%%%%%%% SUPPLEMENTAL INFO %%%%%%%%%%%%%%%%%%%
\newpage

\begin{center}
  \huge {\bf Supplementary Materials}
\end{center}

\setcounter{table}{0}
\setcounter{figure}{0}
\renewcommand{\thetable}{A\arabic{table}}  
\renewcommand{\thefigure}{A\arabic{figure}}

 \begin{table}[H]
 \caption{Data mechanisms used to construct the odds ratio of nonresponse for the PPM model as a function of $\phi$ for fixed $X$.}
 \label{tab:ORMech}
 \resizebox{\columnwidth}{!}{%
 \begin{tabular}{clcccclcc}
 \hline
 \multirow{2}{*}{\begin{tabular}[c]{@{}c@{}}Data\\ Mechanism\end{tabular}} &
   &
 \multicolumn{4}{c}{Respondent Parameters} &
  &
   \multicolumn{2}{c}{Non-respondent Parameters} \\ \cline{3-6} \cline{8-9} 
  &
   &
   $\mu_x^{(1)}$ &
   ${\sigma_x^{(1)}}^2$ &
   $\rho_{xy}^{(1)}$ &
   ${\sigma_{y|x}^{(1)}}^2$ &
  &
  $\mu_x^{(0)}$ &
  ${\sigma_x^{(0)}}^2$ \\ \hline
 1  &  & 1 & 1 & 0.2 & 0.96 &  & 0.8 & 0.90  \\
 2  &  & 1 & 1 & 0.5 & 0.75 &  & 0.8 & 0.90 \\
 3  &  & 1 & 1 & 0.8 & 0.36 &  & 0.8 & 0.90 \\
 4  &  & 1 & 1 & 0.2 & 0.96 &  & 1.2 & 0.90  \\
 5  &  & 1 & 1 & 0.5 & 0.75 &  & 1.2 & 0.90 \\
 6  &  & 1 & 1 & 0.8 & 0.36 &  & 1.2 & 0.90 \\ \hline
 7  &  & 1 & 1 & 0.2 & 0.96 &  & 0.8 & 1.00 \\
 8  &  & 1 & 1 & 0.5 & 0.75 &  & 0.8 & 1.00 \\
 9  &  & 1 & 1 & 0.8 & 0.36 &  & 0.8 & 1.00 \\
 10 &  & 1 & 1 & 0.2 & 0.96 &  & 1.2 & 1.00 \\
 11 &  & 1 & 1 & 0.5 & 0.75 &  & 1.2 & 1.00 \\
12 &  & 1 & 1 & 0.8 & 0.36 &  & 1.2 & 1.00 \\ \hline
 13 &  & 1 & 1 & 0.2 & 0.96 &  & 0.8 & 1.10 \\
14 &  & 1 & 1 & 0.5 & 0.75 &  & 0.8 & 1.10 \\
15 &  & 1 & 1 & 0.8 & 0.36 &  & 0.8 & 1.10 \\
16 &  & 1 & 1 & 0.2 & 0.96 &  & 1.2 & 1.10 \\
17 &  & 1 & 1 & 0.5 & 0.75 &  & 1.2 & 1.10 \\
18 &  & 1 & 1 & 0.8 & 0.36 &  & 1.2 & 1.10 \\ \hline
\end{tabular}%
}
\end{table}

\begin{figure}[H]
\centerline{\includegraphics[scale=1]{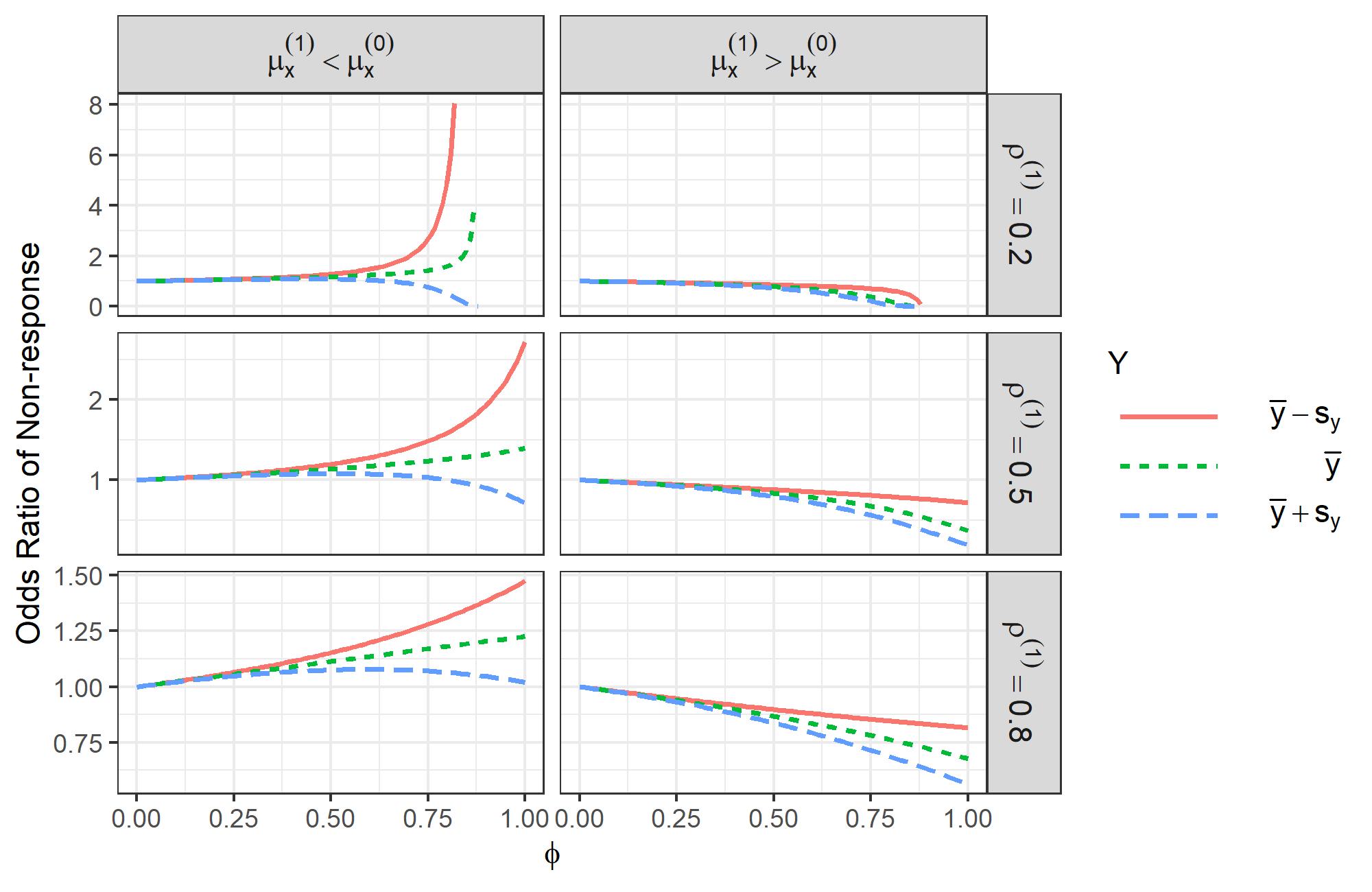}}
\caption{Odds ratio of non-response as a function of $\phi$ when ${\sigma_x^{(1)}}^2 > {\sigma_x^{(0)}}^2$ for different levels of outcome variable, $Y$. The three different levels of the outcome variable were the respondent mean and 1 standard deviation below and above the respondent mean.  The proxy variable was fixed at the overall mean of $X$.}
\label{fig:OR_mech_1_6}
\end{figure}

\begin{figure}[H]
\centerline{\includegraphics[scale=1]{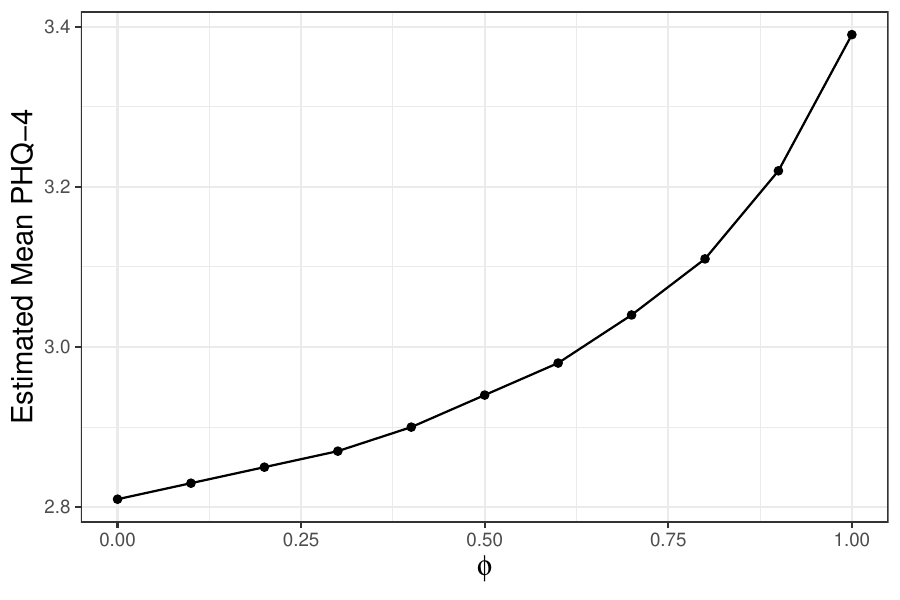}}
\caption{MLE estimates of the (marginal) mean PHQ-4 score under a set of $\phi$ values}
\label{fig:MLE_estimates}
\end{figure}

\end{document}